# Design and Implementation of Hierarchical Visual Cryptography with Expansionless Shares

Pallavi Vijay Chavan[1], Dr. Mohammad Atique[2] and Dr. Latesh Malik[3]

[1]Research Scholar-Dept. of Comp. Sci. & Engg., GHRCE, Nagpur, MH, India
[2]Associate Professor-PG Dept. of Computer Science, SGBAU, Amravati,MH, India
[3] Professor- Dept. of Comp. Sci. & Engg., GHRCE, Nagpur, MH, India

## ABSTRACT

*Novel idea of hierarchical visual cryptography is stated in this paper. The key concept of hierarchical visual cryptography is based upon visual cryptography. Visual cryptography encrypts secret information into two pieces called as shares. These two shares are stacked together by logical XOR operation to reveal the original secret. Hierarchical visual cryptography encrypts the secret in various levels. The encryption in turn is expansionless. The original secret size is retained in the shares at all levels. In this paper secret is encrypted at two different levels. Four shares are generated out of hierarchical visual cryptography. Any three shares are collectively taken to form the key share. All shares generated are meaningless giving no information by visual inspection. Performance analysis is also obtained based upon various categories of secrets. The greying effect is completely removed while revealing the secret Removal of greying effect do not change the meaning of secret.*

## 1. INTRODUCTION

Visual cryptography is the art of encrypting visual information such as handwritten text, images etc. The encryption takes place in such a way that no mathematical computations are required in order to decrypt the secret. The original information to be encrypted is called as secret. After encryption, ciphers are generated and referred as shares. The part of secret in scrambled form is known as *share*. Fundamental idea behind visual cryptography is to share the secret among group of *n* participants. In order to share the secret, it is divided into *n* number of pieces called *shares*. These shares are distributed among the participants. To reveal the original secret, each participant provides his own share. Complete knowledge of *n-1* shares is unable to decrypt the secret. Many visual cryptographic schemes exist. The basic scheme is *2 out of 2* visual cryptography in which the secret is partitioned into exactly two parts. To reveal the secret these two shares must participate. Following figure indicates simple example of 2 out of 2 visual cryptography scheme.

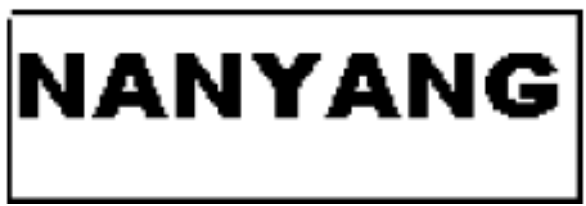

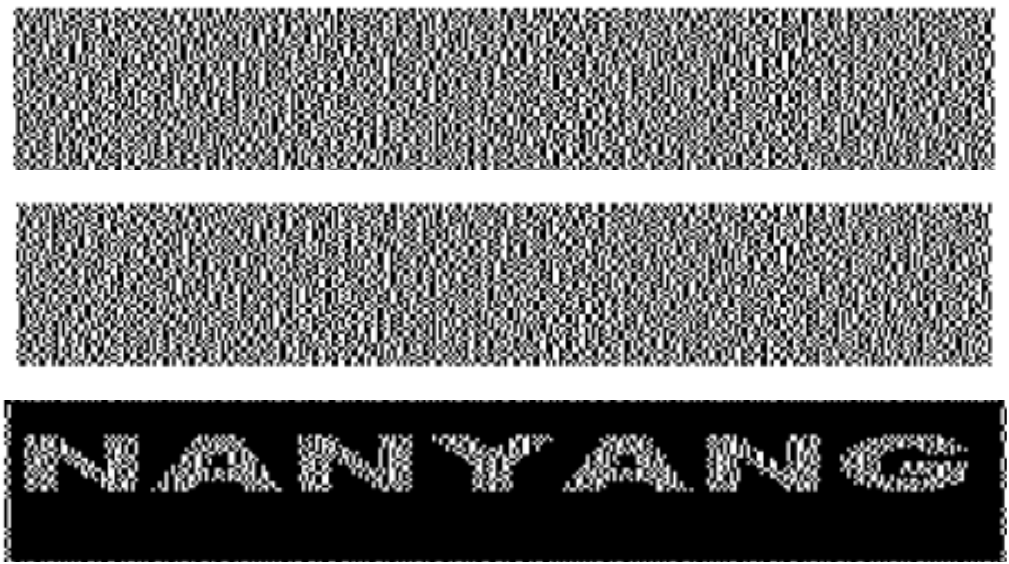

Figure 1. Example of basic visual cryptography

In above figure, graying effect is observed at background. Additional black pixels at the background are forming some pattern and giving raise to greying effect. Due to this effect, meaning of secret remains undisturbed. Hence, greying effect does not have any impact on original secret. Another scheme known as *2* out of *n* scheme, the secret is divided into exactly n shares. To reveal the secret any two shares must participate. Third scheme of VC is formally known as *K* out of *n* scheme in which the secret is divided into exactly *n* parts. To reveal the secret any *K* shares are sufficient. Multiple shares are generated out of visual cryptography. Extended version of third visual cryptography scheme is *n* out of *n* where in secret is divided into *n* shares. All n shares must participate while revealing the secret. Hierarchical visual cryptography abbreviated as HVC is specialization of visual cryptographic schemes. HVC is based upon basic 2 out of 2 visual cryptographic scheme. As the secret is encrypted in multiple levels in HVC, the secrecy of data tends to increase. Now a day's many authentication system exists based upon biometric, passwords but each authentication system is having pitfalls related to the secrecy of data. The idea stated in this paper is further utilized to design the authentication mechanism based on Hierarchical visual cryptography. In this work the size of shares is also reduce giving the expansionless shares. This paper is organized as follows: Section II deals with related work in the area of visual cryptography. Section III describes the key concept of expansionless visual cryptography. In section IV the detailed idea of hierarchical visual cryptography along with experimentation is stated. Experimental results are also represented in this section. The last section V describes conclusion of the overall experimentation.

## 2. RELATED WORK

The concept of visual cryptography formally known as secret sharing was developed by Adi Shamir in 1979. He stated that the secret D is divided in *n* pieces and easily reconstructable from any *k* pieces. He claimed that data is protected after encryption but the key used for encryption could not be protected. He suggested secret sharing to protect the keys used for encryption. According to Shamir, the scheme is *k* out of *n* secret sharing scheme [1]. With the help of basic idea behind visual cryptography, many researchers performed the survey based on various cryptographic schemes. In the survey of VC schemes it has been found that the colorful secret also be shared using visual cryptography [2, 3]. Che Wei Lee used visual cryptography for authentication based on the binary document images with the constraint that the binary document image must follow *portable network graphics* format. In binary documents there are only two possibilities of pixel values viz. black and white. Here authors used 0 to represent black information and 1 to represent white information [4]. Stegenography and visual cryptography concepts were combined by George Abboud to share the hidden message. The idea was quite novel but increased complexity during the computation of shares [5]. In 2008 the concept was proposed to use visual cryptography for banking applications. The authors stated that visual cryptography based signature authentication is more secured than password based authentication. Individual pixel value in the secret is indicated by f(x, y) before processing and by g(x, y) after processing. In any situations for binary secret the values of these two functions are either *0* or *1* [6]. In 2010, cheating prevention mechanism was stated using visual cryptography by Bin Yu [7]. In the same year Li Fang proposed that multiple secrets could be encrypted in the same pairs of shares [8]. Contrast enhancement in visual cryptography was proposed by Thomos Monoth efficiently increased the contrast level of revealed secret [9]. The motivation for Hierarchical visual cryptography oriented from visual cryptography and the need of secured authentication mechanism based on visual cryptography. In 2000, Yang C.n and Laih C.S stated color visual cryptography scheme. According to these authors color information like image is easily encrypted using color VC scheme. Time Complexity was the major issue of discuss in this paper [10]. A segment-based visual cryptography suggested by Borchert [11] can be used only to encrypt the messages containing symbols, especially numbers like bank account number, amount etc. [11]. In 2013, Omprasad and Sonavane developed visual cryptography encoder and decoder by employing the concept of stegenography. Stegenography hides the secret information in such a way that no one information looks like an image. The authors combined visual cryptography with stegenography to obtain better results. The application proposed by authors was Quick Response code [12]. Application of Visual Cryptography to biometric authentication was proposed in 2011 by K. Suryadevara and Naaz. The authors stated that biometric characters are most feasible for authentication. In order to secure the authentication process, the authors applied visual cryptography to biometric character. Every biometric character was acceptable in this technique including fingerprint, iris and tongue. The process of enrollment, verification and identification of biometric character was introduced in this paper. The technique followed 2 out of 2 secret sharing scheme representing 2 out of 2 visual cryptography. The shares generated by this method were expanded version of original secret with expansion ratio of 1:4[13]. In the year 2012, Yi Hui Chen and Pei Yu Lin proposed a novel secret sharing scheme using low computations for authentication. The valuable advantage in this scheme was pixel expansion ratio of 1:1. The secret was reconstructable without loss. The problem identified with this method was authentication with half secret which is infeasible for signature based authentication. The authors used Boolean operations to encrypt the secret [14].

## 3. EXPANSIONLESS VISUAL CRYPTOGRAPHY

In earlier work of visual cryptography, we encrypted the secret with the expansion ratio of *1:4* and later *1:2*. The expansion indicates that if original secret is of size AXB then with expansion ratio *1:4* the shares have size *4AX4B* and with expansion ratio *1:2* the shares reflected are found to be of size *2AX2B*. Due to this expansion hierarchical encryption of secret gets affected. During encryption using HVC initially secret is encrypted using *1:2* expansion ratio giving two shares *S1* and *S2*. If *S1* and *S2* are encrypted with the same expansion ratio independently then the resultant four shares are again expanded forms of *S1* and *S2* (S1 and S2 are already expanded in first level of encryption). This huge expansion in shares affects the space complexity. While superimposing the shares of HVC, the larger transparencies are required. Expansion-less hierarchical visual cryptography is the solution to reduce this expansion in the shares. The requirement of this proposed method is that the secret should be in binary form i.e. black and white passwords, signatures, handwritten text etc. Before encrypting, the original secret is mapped into the size which is multiple of *4*. Before encrypting the secret, it is normalized. The resize function of OCTAVE/MATLAB is chosen to convert any secret whose size is multiple of *4*. After resizing the secret starting with top left corner of secret every *2X2* pixel block is selected for encoding independently. The encoding of *2X2* block is done among various random combinations. If *2X2* block in original secret is entirely black then this block is encrypted using *4* possibilities represented by equation *1 - 8*. There are multiple possibilities for the *2X2* pixel blocks. The encoding combinations for *share1* of entire black block of pixels are represented as below:

$$[1\ 0;\ 0\ 1] \tag{1}$$
$$[0\ 0;\ 1\ 1] \tag{2}$$

$$[1\ 1;\ 0\ 0] \tag{3}$$

$$[0\ 1;\ 1\ 0] \tag{4}$$

The encoding combinations appearing in *share2* are given by:

$$[0\ 1;\ 1\ 0] \tag{5}$$

$$[1\ 1;\ 0\ 0] \tag{6}$$

$$[0\ 0;\ 1\ 1] \tag{7}$$

$$[1\ 0;\ 0\ 1] \tag{8}$$

Similarly, if *2X2* block of secret is found to be entirely white then such block is encrypted with *4* random combinations. *Share1* of encoding for white block of pixels is given by:

$$[1\ 0;\ 0\ 1] \tag{9}$$

$$[0\ 0;\ 1\ 1] \tag{10}$$

$$[1\ 1;\ 0\ 0] \tag{11}$$

$$[0\ 1;\ 1\ 0] \tag{12}$$

Above combinations also represents *share2* as the basic property of encrypting white pixel indicates that the pattern of *share1* and *share2* must be same. Third possibility for *2X2* pixel block is neither pure white nor pure black. In such cases the encryption is done using following combinations. *share1* for random combinations is given below:

$$[0\ 1;\ 1\ 0] \tag{13}$$

$$[1\ 0;\ 1\ 0] \tag{14}$$

$$[0\ 1;\ 1\ 1] \tag{15}$$

$$[1\ 0;\ 1\ 1] \tag{16}$$

Corresponding *share2* combinations are represented by following matrices:

$$[0\ 1;\ 0\ 1] \tag{17}$$

$$[1\ 1;\ 0\ 1] \tag{18}$$

$$[1\ 0;\ 1\ 0] \tag{19}$$

$$[0\ 1;\ 0\ 1] \tag{20}$$

## 3.1. Expansionless Visual Cryptography Algorithm

Input: Secret in binary form
Output: Meaningless shares

1. Read the secret.
2. Covert secret size $AXB$ to exact multiple of *4*.
3. Determine the size of converted secret: *[s1 s2]< ¡*size(secret)
4. For each *i* in *s1* with step size *2*

      For each *j* in *s2* with step size *2*
      if(*2X2* block is entirely black)

      Randomly select the pattern from equation *1-4* to generate *share1* and from equation *5-8* to generate *share2*.

      Else
      if(*2X2* block is entirely white)
      Randomly select the pattern from equation *9-12* to genrate *share1* and *share2*
      Else
      Encrypt the block using random selection among equations *13-16* to generate *share1* and using equations *17-20* to generate share2.
      *End for*
*End for*
5. Exit.

## 4. HIERARCHICAL VISUAL CRYPTOGRAPHY EXPERIMENTATION

Hierarchical visual cryptography encrypts the secret in two levels. Initially the secret is encrypted in two different meaningless shares called *share1* and *share2*. This is the first level of Hierarchical visual cryptography. In the second level, these two shares are encrypted independently. *Share1* is encrypted to yield *share11* and *share12*. Similarly, *share2* is encrypted to yield *share21* and *share22*. At the end of second level of HVC four shares are available. The key share formation takes place here. Among these four shares any three shares are taken and mapped to yield the *key share*. Table 1 shows the mapping of key share. With the help of mapping stated here, the *key share* found to have blacker tendency.

Table 1. Key Share Generation

| Share12 | Share21 | Share22 | Key Share |
|---------|---------|---------|-----------|
| 0 | 0 | 0 | 1 |
| 0 | 0 | 1 | 1 |
| 0 | 1 | 0 | 0 |
| 0 | 1 | 1 | 0 |
| 1 | 0 | 0 | 0 |
| 1 | 0 | 1 | 1 |

| 1 | 1 | 0 | 0 |
|---|---|---|---|
| 1 | 1 | 1 | 0 |

Another mapping method is stated here with soft computing approach. While mapping these three shares to key share, following inference rules are applied. These inference rules are defining the intelligence of an algorithm. The concentration of black and white pixels in key share is determined based upon concentrations of three input shares. These if-then rules are applicable to each pixel of the *share12, share21, share22*.

1. If (*share12, share21, share22* is *black)* then *keyshare* pixel is black;
2. If (*share12, share21* is black and *share22* is white) then *keyshare* pixel is black;
3. If (*share12, share22* is black and *share21* have white) then *keyshare* pixel is black;
4. If (*share12* is black and *share21, share22* is white) then *keyshare* pixel is black;
5. If (*share12* is white and *share21, share22* is black) then *keyshare* pixel is black;
6. If (*share12, share22* is white and *share21* is black) then *keyshare* pixel is white
7. If (*share12, share21* is white and *share22* is black) then *keyshare* pixel is black;
8. If (*share12, share21, share22* is white) then *keyshare* pixel is black.

## 4.1. Methodology for Hierarchical Visual Cryptography

Hierarchical visual cryptography is defined on basis of visual cryptography. Simple visual cryptography divides original secret in two parts. Each part is known as share. To reconstruct the secret, both shares are stacked together. This technique is known as 2 out of 2 visual cryptography. Hierarchical visual cryptography also encrypts the secret information in two shares at the first stage. Later, these two shares are encrypted individually to generate subsequent shares. This is the second level of hierarchy in HVC. At the end, four resultant shares are found. Out of these four shares, any three shares are taken to generate the key share. This stage is identified as third level of HVC. Finally, HVC scheme gives two resultant shares out of which one is handed over to the user for authentication and another share is along with database. The shares are generated using Naor and Shamir scheme of two out of two visual cryptography. Algorithm for Encrypting Secret using hierarchical visual cryptography is given below:

1. Begin
2. Read original secret.
3. Resize the secret to standard size.
4. Encrypt resized secret using 2 out of 2 VC scheme.
    *Share 1* and *Share 2* are generated here.
5. *Share 1* is encrypted using 2 out of 2 VC.
    *Share 11* and *share 12* are generated here.
6. *Share 2* is encrypted separately using 2 out of 2 VC.
    *Share 21* and *share 22* are generated here.
7. *Share 11*, *Share 12* and *Share 21* are combined to form *key share*
    (Inference rules are defined to generate key share pixel value.)
8. Output remaining share and *key share.*
9. End.

## 4.2. Experimental Results

The experimentation with expansionless hierarchical visual cryptography is done with reference to the following secret shown in f*igure 2*. Above secret is encrypted using expansionless visual cryptography. After encryption, *share1* and *share2* are generated at the first level of *HVC* which are represented by fi*gure 3*. Encryption at the second level is shown in fig*ure 4*. The key share is shown in fi*gure 5*. Finally *figure 6* indicates revealed secret.

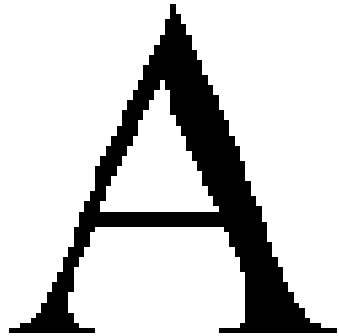

Figure 2. Original secret

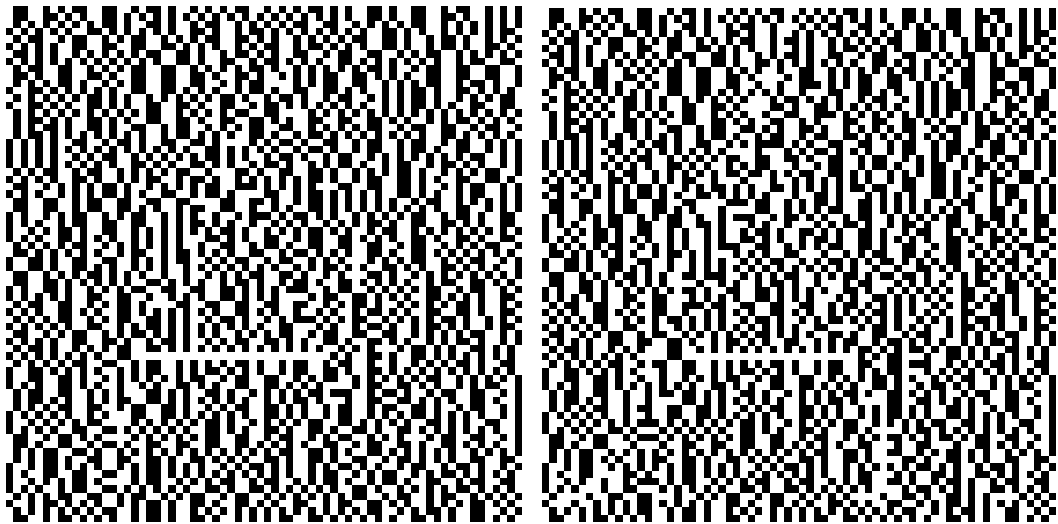

Figure 3. Shares at first level of hierarchy

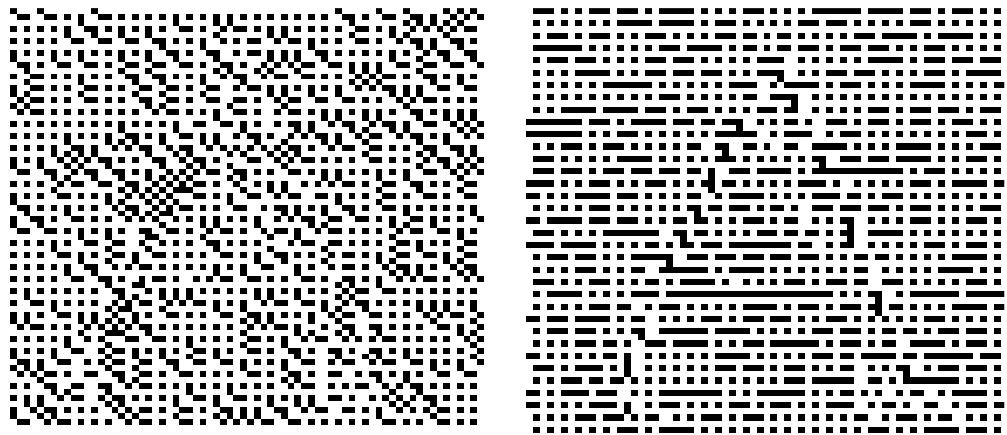

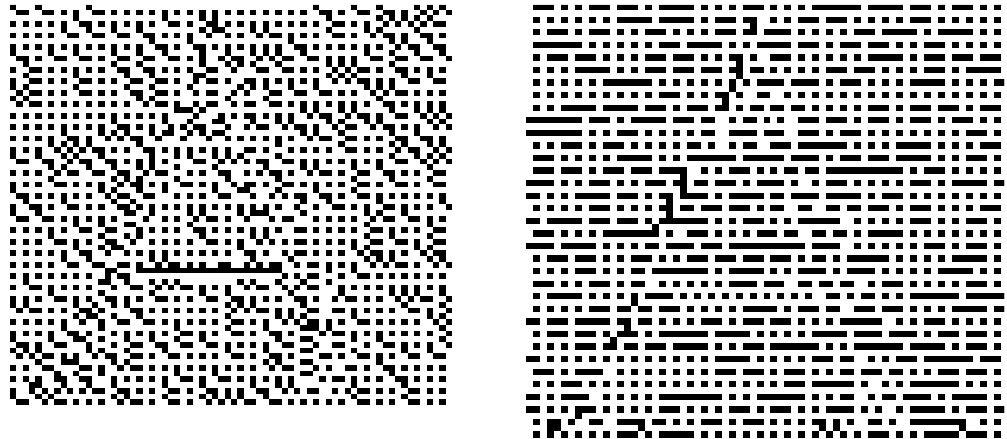

Figure 4. Shares at second level of hierarchy

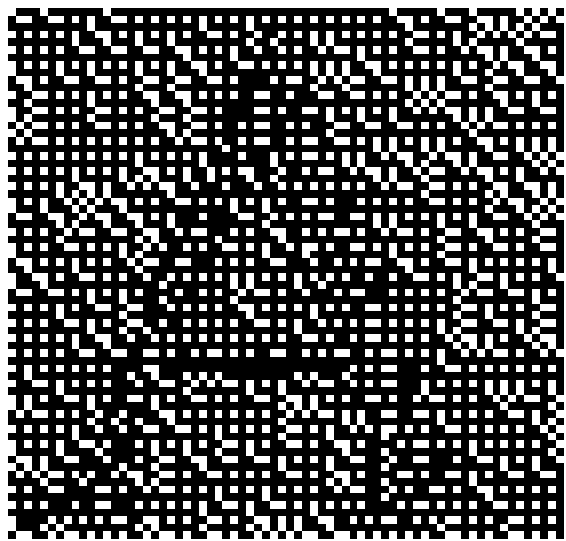

Figure 5. Key share

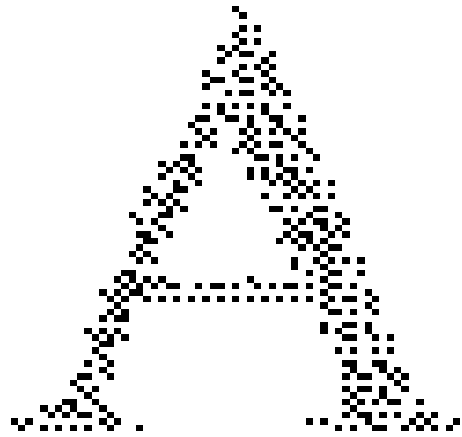

Figure 6. Revealed secret

## 4.3. Experimental Analysis

Concentration of black and while pixels in the shares and revealed secret are analyzed. Each entity involved in hierarchical visual cryptography represents different concentration of black and white information. As the secret changes over time, corresponding concentration varies with respect secret. Percentage of black and white pixels remains constant for different category of secrets like images, handwritten text and textual passwords. Following Table denotes exact concentration of black and white pixels in shares and secret. It has been analyzed that the concentration of white pixels decreases drastically in first level of hierarchical visual cryptography. This concentration remains steady for second level and later increases at the end of hierarchy. Key share reflects high concentration of black pixels. The revealed secret represents high concentration of white pixels. High concentration in reconstructed secret is helpful to detect the secret easily during authentication. It leads to removal of graying effect observed in basic Shamir scheme.

Table 2. Statistics of pixels in HVC encoder

| Entity | No. of black pixels | No. of white pixels | Total pixels |
|---|---|---|---|
| Secret | 2364 | 6744 | 9108 |
| Resized Secret | 4989 | 15159 | 20148 |
| Share1 | 10479 | 9669 | 20148 |
| Share2 | 10463 | 9685 | 20148 |
| Share11 | 10479 | 9669 | 20148 |
| Share12 | 10463 | 9685 | 20148 |
| Share21 | 6716 | 13484 | 20148 |
| Share22 | 6716 | 13432 | 20148 |
| Revealed secret | 2024 | 18124 | 20148 |

Following graph indicates white pixel concentration in encrypting entity and corresponding shares. The statistics of hierarchical visual cryptography encoder are mapped in this graph.

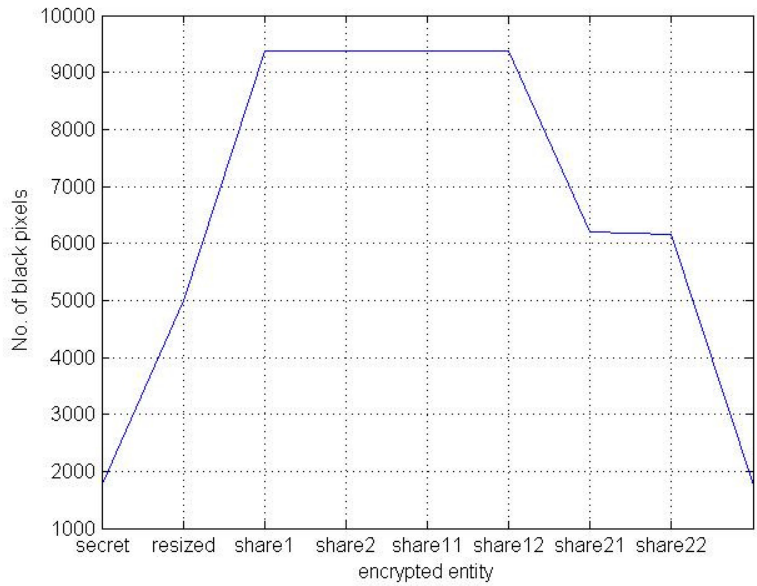

Figure 7. Representation of black pixels in entities

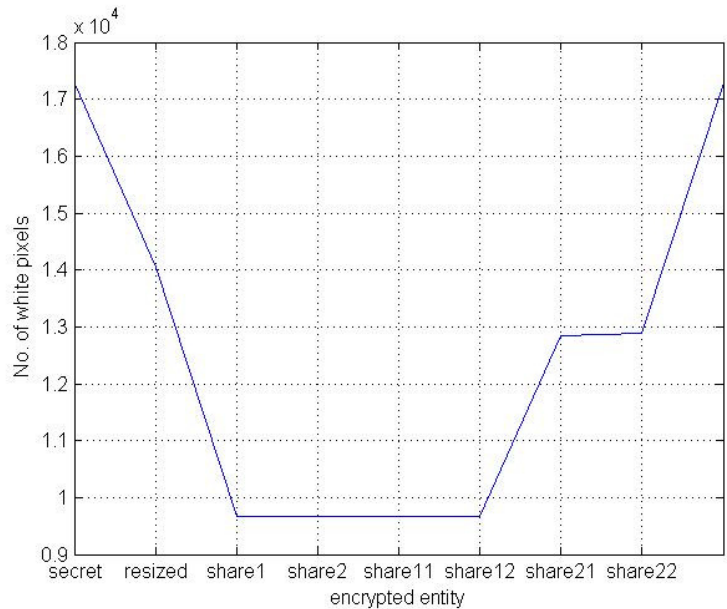

Figure 7. Representation of white pixels in entities

## 5. CONCLUSIONS

In this paper, we proposed and experimented hierarchical visual cryptography. Related work in area of visual cryptography is also discussed in this paper. Encryption at each level of HVC is expansionless. A share generated out of HVC represents the same size of secret. The *keyshare* generated is having random nature. It has been observed that the expansionless shares consume less memory. Graying effect is reduced to zero. In earlier work of visual cryptography it has been observed that expansion of secret taking place after encryption. Thus reflects some greying effect.

The authentication mechanism based on hierarchical visual cryptography is identified as a future scope of this work. The authors encrypted various categories of secrets to compare the results.

## ACKNOWLEDGEMENTS

The authors are thankful to Shamir for providing basic platform of visual cryptography. We are also thankful to all the authors those who contributed in the area of visual cryptography.

**Authors**

Ms. Pallavi Vijay Chavan is a research scholar at G. H. Raisoni College of Engineering, Higna, Nagpur, MH, India and working as an assistant professor (Sr.Gr.) at Bapurao Deshmukh College of Engineering, Sevagram, MH, India. She did her BE in Computer Engineering in 2003 from Nagpur University, Nagpur and ME in Computer Science and Engineering from SRTMNU, Nanded. She is having 9 years of teaching experience. She has published 10 research papers in International conference s, 4 research papers in International journals and 6 papers in national conferences. She is a  life  member of ISTE. 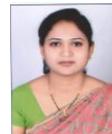

Dr. Mohammad Atique is working as an associate professor in P.G. Department of compu ter science at Sant  Gadge Baba Amravati University Amravati, MH, India. He has published many research papers in International journals and conferences as well as in national journals and conferences. He is member of many professional bodies like ISTE. His area of interest is soft computing and operating system. He is recipient of research grants from research agencies like AICTE and UGC. 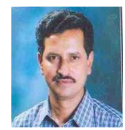

Dr. Latesh Malik is professor and head of computer science and engineering department at G. H. Raisoni College of Engineering, Nagpur. She has published many research papers in the area of pattern recognition. Her area of interest is Handwritten text recognition. 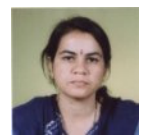